\documentclass[11pt]{amsart}
\usepackage{amssymb,amsmath,amscd,amsthm,amsfonts,wasysym,mathrsfs,amsxtra}
\input xy
\xyoption{all}

\newcommand{\OO}{\mathscr{O}}

\newcommand{\arinj}{\ar@{^{(}->}}
\newcommand{\arsurj}{\ar@{->>}}

\newtheorem{theorem}{Theorem}[section]

\newtheorem{conjecture}[theorem]{Conjecture}

\theoremstyle{definition}

\theoremstyle{remark}

\numberwithin{equation}{section}

\begin{document}
\title[Higher Rank Refined Donaldson-Thomas Invariants]{On Some Computations of Higher Rank Refined Donaldson-Thomas Invariants}

\author[Wu-yen Chuang]{Wu-yen Chuang}
\address{Department of Mathematics, National Taiwan University, Taipei, Taiwan} \email{wychuang@ntu.edu.tw}
\author[Chien-Hsun Wang]{Chien-Hsun Wang}
\address{Department of Physics, National Taiwan University, Taipei, Taiwan} \email{chwangk@gmail.com}


\begin{abstract}
We present some computations of higher rank refined Donaldson-Thomas invariants on local curve geometries, corresponding to local D6-D2-D0 or D4-D2-D0 configurations. 
A refined wall-crossing formula for invariants with higher D6 or D4 ranks is derived and verified to agree with the existing formulas under the unrefined limit. Using the formula,
refined invariants on the $(-1,-1)$ and $(-2,0)$ local rational curve with higher D6 or D4 ranks are computed. 
\end{abstract}

\maketitle


\section{Introduction}
Recently BPS state counting has become an active research area in superstring and supersymmetric gauge theory. A complete understanding of the BPS spectra of a theory would be crucial towards a non-perturbative formulation of the theory under investigation.  

In this paper we study the refined BPS state counting, which has an M-theoretical origin. Consider an M-theory compactification on a smooth projective Calabi-Yau threefold $X$. M2-branes wrapping holomorphic curves in $X$ yield BPS particle states in the five dimensional effective theory. These particles are electrically charged under the low energy $U(1)$ gauge fields, and the charge lattice is naturally identified with the second homology lattice $H_2(X,\mathbb{Z})$. Quantum states of massive particles in five dimensions also form multiplets of its little group $SU(2)_L\times SU(2)_R\subset SO(4,1)$. The irreducible representations of $SU(2)_L\times SU(2)_R$ could be labeled by pairs of half-integers $(j_L,j_R)\in \big({1\over 2}\mathbb{Z}\big)^2$, which are the left and right moving spin quantum numbers. In other words, the space of five dimensional BPS states admits the following direct sum decomposition
\[
\mathcal{H}_{BPS}(X)\simeq \bigoplus_{\beta\in H_2(Y,\mathbb{Z})} 
\bigoplus_{j_L,j_R\in {1\over 2}\mathbb{Z}} \mathcal{H}_{BPS}(X,\beta,j_L,j_R) , 
\] which is the origin of the refined BPS invariants. 
The refined  Gopakumar-Vafa invariants are the BPS degeneracies 
\[
N(X,\beta,j_L,j_R) =\text{dim}\, \mathcal{H}_{BPS}(X,\beta,j_L,j_R).
\]
The unrefined invariants are BPS indices, 
\[
N(X,\beta,j_L) = \sum_{j_R\in {1\over 2}\mathbb{Z}} 
(-1)^{2j_R+1} (2j_R+1)N(X,\beta,j_L,j_R) . 
\] 

Therefore the unrefined BPS state counting corresponds to the state counting in the BPS Hilbert space $\mathcal{H}_{BPS}(X)$ with a supertrace over the spin quantum number $j_R$, while the refined invariants keep all the spin information. Note that the refined BPS indices are only well-defined when the Calabi-Yau has no complex structure deformation. For a certain class of non-compact Calabi-Yaus, there admits field theory limits such that the gravity decouples in the lower energy. In these cases the BPS states are actually the BPS states in the low energy supersymmetric gauge theory. 

On the other hand, some string theory arguments \cite{Gopakumar:1998ii}\cite{Gopakumar:1998jq} suggest that BPS states should be identified with cohomology classes of moduli spaces of certain stable sheaves on $X$. More precisely, let $\mathcal{M}(X,\beta,n)$ be the moduli space of slope (semi)stable pure dimension one sheaves $F$ on $X$ with numerical invariants 
\[
\text{ch}_2(F)=\beta, \qquad \chi(F)=n, 
\] where $\beta$ is the curve class representing the support of the sheaf. If $\mathcal{M}(X,\beta,n)$ is smooth, the BPS states are in one-to-one correspondence with the cohomology classes of the sheaf moduli. However generically the moduli are singular and mathematically the BPS state counting is believed to become a virtual count of the singular space, involving an integration of one over the virtual cycles defined in \cite{BF} \cite{LT}.

Moreover an interesting fact about the BPS indices is that they are only locally constant and can jump across the walls of marginal stability where the central charge of the states align, such that some BPS states annihilate or some new BPS states emerge. The discrete change is determined by the wall-crossing formula. Mathematically wall-crossing formulas are derived by the unique Harder-Narasinham filtration of an object with respect to a stability condition and the algebra of the constructible functions on the moduli \cite{ReinekeQ}. 

Since it is difficult to deal with the virtual cycles directly, one usually apply virtual torus localization \cite{GP} and wall-crossing formulas by Joyce-Song \cite{genDTI} or Kontsevich-Soilbelman \cite{Kontsevich:2008fj} to obtain the explicit results. 

In this paper we actually study D6-D2-D0 and D4-D2-D0 configurations in string theory.
The M-theoretical origin of the rank one D6-D2-D0 states is given in \cite{Dijkgraaf:2006um}. The multiple D6 configuration can be obtained
from M-theory via a multi-centered Taub-NUT space. The D4-D2-D0 states, on the other hand, comes from an M5-M2 configuration. In both cases, upon the compactification along the M-theory circle, the 5d spin quantum numbers $(j_L,j_R)\in \big({1\over 2}\mathbb{Z}\big)^2$ will get related to the D0-brane charge and the spin quantum number $s$ of the 4d theory \cite{Gaiotto:2005gf}. Conjecturally the GV moduli space admits a fibration structure such that there exist two Lefschetz actions along both the fiber and the base. And the diagonal combination of the two is the usual Lefschetz action on the moduli space \cite{Katz:1999xq}, whose representation gives the 4d spin quantum number $s$.
   
The rank one wall-crossing formula is physically derived in \cite{Denef:2007vg} and its refined generalization is done in \cite{Dimofte:2010wxa}. The purpose of the present paper is to study higher rank refined invariants on the local curve geometries. First we recall some existing results in the unrefined cases. In the unrefined cases, if the generating function of the invariants is known in a certain chamber, successive applications of the wall-crossing formulas will, in principle, give us the generating functions of all the other chambers on the moduli space. The higher rank generalization with multiple D6 branes on local curve geometries is done in \cite{Chuang:2010wx}, in which the multicover formula and stacky invariants of the strictly semistable objects are studied. The higher rank invariants in projective Calabi-Yau 3-folds are studied in \cite{ranktwo} and \cite{ranktwoGW}. The generating function of multiple D4 branes on conifold has also been computed \cite{Nishinaka:2010fh}.

Physically local invariants with multiple D6 or D4 branes are also interesting in string theory. Such invariants are responsible for certain subleading corrections in the OSV conjecture \cite{Denef:2007vg}\cite{Ooguri:2004zv}. It would be interesting to investigate the implications of our results in this famous conjecture. We leave it as a future project.

The paper is organized as follows. In the next section we will derive the rank two refined wall-crossing formula, applying Kontsevich-Soibelman's motivic wall-crossing formula. In section 3 and 4 we apply it to compute the generating function of local rational curve for two D6 and D4 branes respectively. In section 5 we discuss some recent results of the higher rank invariants and in the last section we conclude the paper.

\medskip
\noindent
{\it Acknowledgements.} WYC and CHW were supported by NSC grant 101-2628-M-002-003-MY4. 
WYC would like to thank Emanuel Diaconescu and Guang Pan for the related collaboration in the past few years.
CHW would like to thank T. Nishinaka for useful discussions.

\section{Refined rank two wall-crossing formula}

Let $\Sigma_g$ be a projective curve of genus $g$ over $\mathbb{C}$. Then the total space of $\mathcal{O}(p)\oplus\mathcal{O}(2g-2-p)$ over $\Sigma_g$ is a non-compact Calabi-Yau threefold with trivial canonical bundle. 
Motivated by string theoretical consideration, ADHM sheaf theory was first introduced by Diaconescu \cite{Diaconescu:2008ct} and the theory has a natural variation of the stability conditions \cite{Diaconescu:2009zf, Chuang:2009pv}. In an asymptotic chamber of the stability condition space the ADHM sheaf theory defined on $\Sigma_g$ is equivalent to admissible pair theory on the projective plane bundle over $\Sigma_g$.
When certain twisting data $(M_1,M_2)$ in the ADHM sheaf theory are chosen such that $M_1^{-1} \simeq \mathcal{O}(p)$ and $M_2^{-1} \simeq \mathcal{O}(2g-2-p)$, this pair theory becomes a stable pair theory on the total space of CY local curve geometry. The key ingredients of the construction consist of a relative version of Beilinson spectral sequence and Fourier-Mukai transformation, in which we use the structure sheaf of the diagonal as Fourier-Mukai kernel.  

In the following we generalize the rank two wall-crossing formula to the refined case in the ADHM sheaf theory on curves. We first use the notation $\gamma=(e,r,w,v) \in \mathbb{Z}^{\times 4}$ to denote the D0, D2, D4 and D6 brane charges respectively. The refined invariants for the charge $\gamma$ and spin $s$ is denoted by $\Omega_s(e,r,w,v)\in\mathbb{Z}$. More precisely the support of $\gamma$ is in $\mathbb{Z}\times\mathbb{Z}_{\geq1}\times\mathbb{Z}_{\geq0}\times\mathbb{Z}_{\geq0}$, due to the construction of the ADHM sheaf theory. We denote the refined invariants of the two sides of a wall $\Omega_s(\gamma)_{\pm}$. Then the integer refined Donaldson-Thomas invariant $\overline{\mathrm{DT}}(\gamma;y)\in\mathbb{Z}(y)$ is defined by \cite{Kontsevich:2008fj,Dimofte:2009bv,Dimofte:2009tm,Manschot:2010qz}
\begin{equation}
\overline{\mathrm{DT}}(\gamma;y)=\sum_{s}(-y)^{s}\Omega_{s}(\gamma) 
\end{equation} which is an integral coefficient Laurent polynomial in $y$. We also define the rational refined Donaldson-Thomas invariants by the refined multicover formula \cite{Chuang:2010ii}
\begin{equation} \label{3}
\mathrm{DT}(\gamma;y)=\sum_{k|\gamma}\frac{1}{k[k]_{y}}\overline{\mathrm{DT}}\big(\frac{\gamma}{k};y^k\big) , 
\end{equation}
where the quantum integer $[n]_{y}$ is defined by
\begin{align*}
[n]_{y}=\frac{y^n-y^{-n}}{y-y^{-1}} . 
\end{align*}
To describe the refined wall-crossing formula, we introduce the infinite dimensional Lie algebra spanned by the generators ${\hat{e}_{\gamma}}$ for each $\gamma$ such that
\begin{equation}\label{alg}
\hat{e}_{\gamma'}\hat{e}_{\gamma''}=(-y)^{\langle\gamma',\gamma''\rangle}\hat{e}_{\gamma'+\gamma''}
\end{equation}
with the Lie bracket
\begin{align}\label{bracket}
&[\hat{e}_{\gamma'},\hat{e}_{\gamma''}]=\kappa(\langle\gamma',\gamma''\rangle)\hat{e}_{\gamma'+\gamma''}.
\end{align} 
Where $\kappa(x)=(-y)^{x}-(-y)^{-x}$ and $\langle,\rangle$ is the Euler form defined by $\langle \gamma_1,\gamma_2\rangle = \sum_i (-)^i \text{dim} \text{Ext}^{i}(\gamma_1,\gamma_2).$ In the following of this section, we denote the charges $\gamma=(\alpha,v)$, where  $\alpha=(r,e)\in\mathbb{Z}_{\geq1}\times\mathbb{Z}$ represents the D2, D0 brane charge and $v\in\mathbb{Z}_{\geq 0}$ represents the D6 or D4 brane charge, since we do not consider the case with both D6 or D4 brane charges. 
In order to derive the rank two refined wall-crossing formula, we truncate the Lie algebra to D6/D4 brane charge being equal to or less than two, using the following notation: 
\begin{align*}
& [\hat{e}_{(\alpha_1,v_1)}, \hat{e}_{(\alpha_2,v_2)}]_{\leq 2} = 
\left\{\begin{array}{ll} 
[\hat{e}_{(\alpha_1,v_1)}, \hat{e}_{(\alpha_2,v_2)}] & \quad \mathrm{if}\ v_1+v_2\leq 2, \\
0 & \quad \mathrm{otherwise}.\\
\end{array}\right. \\
\end{align*}
In the unrefiend limit $y\rightarrow1$, we have $e_{\gamma}:=\lim\limits_{y\to1}\hat{e}_{\gamma}(y^2-1)^{-1}$ and 
\begin{equation*}
\lim\limits_{y\to1}(y^2-1)^{-1}\big((-y)^{\langle\gamma',\gamma''\rangle}-(-y)^{-\langle\gamma',\gamma''\rangle}\big)=(-1)^{\langle\gamma',\gamma''\rangle}\langle\gamma',\gamma''\rangle
\end{equation*} 
and the refined Donaldson-Thomas invariants $\mathrm{DT}(\gamma;y)$ reduce to the numerical Donaldson-Thomas $\mathrm{DT}(\gamma)$ invariants.

To define the stability condition we first introduce the $\mu$ slope function $\mu(\gamma)=e/r.$
Then the $\delta$ slope for the charge $\gamma=(\alpha,v)$ is defined to be
\begin{equation*}
\mu_{\delta}(\gamma)=\mu(\gamma)+\frac{v\delta}{r},
\end{equation*}
where the stability parameter $\delta \in \mathbb{R}$ is first introduced in \cite{Diaconescu:2009zf}. When $\delta$ is asymptotically  large, the invariants computed in this stability chamber corresponds the stable pair theory defined by Pandharipande and Thomas \cite{stabpairs-I}.
Therefore we will call it PT chamber for short. 

For any two pairs $\alpha =(r_{\alpha}, e_{\alpha})$ and $\beta =(r_{\beta}, e_{\beta})$
The critical stability parameter $\delta_c$ of type $(r,e,2)$, $(r,e)\in \mathbb{Z}_{\geq 1}\times \mathbb{Z}$ is
\begin{equation}
\frac{e_{\alpha} + \delta_c}{r_{\alpha}}=\frac{e_{\beta}}{r_{\beta}}= \mu_{\delta_c}(\gamma),
\end{equation}
so that any $\eta\in \mathbb{Z}_{\geq 1}\times \mathbb{Z}$ 
with $\mu_{\delta_c}(\eta) = \mu_{\delta_c}(\gamma)$ can 
be uniquely written as $\eta = (q\beta,0), (\alpha+ q\beta,1),$ or 
$(2\alpha+ q\beta,2)$, with $q\in \mathbb{Z}_{\geq 0}$.

The Euler form of two charges both with zero D6 brane charge $\gamma=(\alpha,0)$ vanishes so the Lie algebra of the generators in this case become $\hat{e}_{\gamma'}\hat{e}_{\gamma''}=\hat{e}_{\gamma'+\gamma''}$. Therefore we have $(\hat{e}_{\gamma})^{k}=\hat{e}_{kn}$. 
For convenience we introduce the following notations: $\hat{e}_{\alpha}=\hat{e}_{(\alpha,0)}$, $\hat{f}_{\alpha}=\hat{e}_{(\alpha,1)}$ and $\hat{g}_{\alpha}=\hat{e}_{(\alpha,2)}$.

Now we introduce the quantum dilogarithm
\begin{equation*}
\mathbf{E}(x)=\prod_{i=0}^{\infty}(1-(-y)^{2i+1}x)^{-1}=\exp\bigg(\sum_{k=1}^{\infty}\frac{x^k}{k(y^{k}-y^{-k})}\bigg) .
\end{equation*}
Then we list all the relevant wall-crossing factors appearing in the refined Kontsevich-Soibleman wall-crossing formula as follows: 
\begin{align*}
\prod_{q\geq 0 }\hat{U}_{\hat{e}_{q\beta}}&=\prod_{q\geq 0 }\prod_{s}\mathbf{E}(y^{s}\hat{e}_{q\beta})^{-(-1)^{s}\Omega_{s}(q\beta)}
=\exp\bigg(\sum_{q\geq 0 }\sum_{k=1}^{\infty}\frac{\overline{\mathrm{DT}}(q\beta;y^k)}{k(y^k-y^{-k})}\hat{e}_{kq\beta}\bigg)\\&=\exp\bigg(\sum_{q\geq 0 }\mathrm{DT}(q\beta;y)\hat{e}_{q\beta}\bigg),
\\\nonumber
\hat{U}^{\pm}_{\hat{f}_{\alpha+q\beta}}&=\prod_{s}\mathbf{E}(y^{s}(\hat{f}_{\alpha+q\beta}))^{-(-1)^{s}\Omega_{s}^{\pm}(\alpha+q\beta,1)}
\\\nonumber
&=\exp\bigg(\frac{1}{y-y^{-1}}\overline{\mathrm{DT}}_{\pm}(\alpha+q\beta,1;-y)\hat{f}_{\alpha+q\beta}\\&+\frac{1}{y-y^{-1}}\overline{\mathrm{DT}}_{\pm}(\alpha+q\beta,1;y^2)\frac{1}{2[2]_y}\hat{g}_{2\alpha+2q\beta}\bigg),
\\\nonumber
\hat{U}^{\pm}_{\hat{g}_{2\alpha+q\beta}}&=\prod_{s}\mathbf{E}(y^{s}(\hat{g}_{2\alpha+q\beta}))^{-(-1)^{s}\Omega_{s}^{\pm}(2\alpha+q\beta,2)}\\ \nonumber
&=\exp\bigg(\frac{1}{y-y^{-1}}\overline{\mathrm{DT}}_{\pm}(2\alpha+q\beta,2;y)\hat{g}_{2\alpha+q\beta}\bigg).
\end{align*}
Note that the invariants involving only the D2/D0 charges are insensitive to the variation of the stability parameter $\delta$, while all other invariants have $\pm$ superscripts.

Let $\delta_c$ be a critical value of $\delta$. The Kontsevich-Soibelman wall-crossing formula states that the product of quantum symplectomorphisms with increasing $\mu_{\delta}$ slopes do not change under wall-crossing of the value $\delta_c$, which gives the following identity: 
\begin{align}\label{01}
&\prod_{q\geq 0 }\hat{U}_{\hat{e}_{q\beta}}\prod_{q\geq 0,\downarrow q }\hat{U}^{+}_{\hat{g}_{2\alpha+q\beta}}\prod_{q\geq 0,\downarrow q}\hat{U}^+_{\hat{f}_{\alpha+q\beta}}=\prod_{q\geq 0,\uparrow\downarrow q }\hat{U}^{-}_{\hat{f}_{\alpha+q\beta}}\prod_{q\geq 0,\uparrow q }\hat{U}^{-}_{\hat{g}_{2\alpha+q\beta}}\prod_{q\geq 0 }\hat{U}_{\hat{e}_{q\beta}} .
\end{align}
By $\eqref{3}$, the refined multicover formula in this case yields
\begin{equation} \label{refinemc} \mathrm{DT}_{\pm}(2\alpha+q\beta,2;y)=\overline{\mathrm{DT}}_{\pm}(2\alpha+q\beta,2;y)+\frac{1}{2[2]_{y}}\overline{\mathrm{DT}}_{\pm}(\alpha+q\beta/2,1;y^2),
\end{equation}
Expanding both side of \eqref{01}, omitting terms involving $\hat{f_{\gamma}}$ which contribute only to rank one formula and applying \eqref{refinemc}, the equation \eqref{01} yields

\begin{align} \label{wcf1}
&\exp\bigg(\frac{1}{y-y^{-1}}\sum_{q\geq 0 }\mathrm{DT}_{-}(2\alpha+q\beta,2;y)\hat{g}_{2\alpha+q\beta}\\ \nonumber &+\frac{1}{2(y-y^{-1})^2}\sum_{q_2>q_1\geq0}\kappa(\chi(q_1\beta,q_2\beta))\overline{\mathrm{DT}}_{-}(\alpha+q_1\beta,1;y)\overline{\mathrm{DT}}_{-}(\alpha+q_2\beta,1;y)\hat{g}_{2\alpha+2(q_1+q_2)\beta}\bigg)\\ \nonumber&=\exp\big(\sum_{q\geq 0}\mathrm{DT}(q\beta;y)\hat{e}_{q\beta}\big)\exp\bigg(\sum_{q\geq 0 }\frac{\mathrm{DT}_{+}(2\alpha+q\beta,2;y)}{y-y^{-1}}\hat{g}_{2\alpha+q\beta}\\  \nonumber &+\sum_{q_2>q_1\geq0}\frac{\kappa(\chi(q_1\beta,q_2\beta))}{2(y-y^{-1})^2}\overline{\mathrm{DT}}_{+}(\alpha+q_1\beta,1;y)\overline{\mathrm{DT}}_{+}(\alpha+q_2\beta,1;y)\hat{g}_{2\alpha+2(q_1+q_2)\beta}\bigg)\\ \nonumber
&\exp\big(-\sum_{q\geq 0}\mathrm{DT}(q\beta;y)\hat{e}_{q\beta}\big) \ .
\end{align}
Applying Baker-Campell-Hausdorff (BCH) formula
\begin{equation*}
\begin{aligned}
\text{exp}(A) \text{exp}(B) \text{exp} (-A) & = \text{exp}( \sum_{n=0} \frac{1}{n!} (Ad(A))^n B )\\ & 
= \text{exp} ( B + [A,B] +\frac{1}{2} [A,[A,B]]+ \cdots),\\
\end{aligned}
\end{equation*}
to the RHS of \eqref{wcf1} yields
\begin{align*}
&\text{RHS}=\exp\bigg(\sum_{ \substack{q\geq0, l\geq 0 \\ q_i>0}}
\frac{1}{y-y^{-1}}\mathrm{DT}_{+}(2\alpha+q\beta,2;y)\times \\ \nonumber &\frac{1}{l!}\prod_{i=1}^{l}(\frac{1}{y-y^{-1}})^{i}\kappa(f_2(q_{i}\beta))\mathrm{DT}(q\beta;y)\hat{g}_{2\alpha+(q+q_{1}+..q_{l})\beta}\\&+\sum_{\substack{q_1' > q_2'\geq 0\\ l\geq0, q_i>0}}\frac{1}{2(y-y^{-1})^2}\sum_{l}\kappa(g(q_1\beta,q_2\beta))\overline{\mathrm{DT}}_{+}(\alpha+q_1\beta,1;y)\overline{\mathrm{DT}}_{+}(\alpha+q_2\beta,1;y)\\&\times\prod_{i=1}^{l}(\frac{1}{y-y^{-1}})^{i}\kappa(f_2(q_{i}\beta))\mathrm{DT}(q\beta;y)\hat{g}_{2\alpha+(q'_{1}+q'_{2}+q_{1}+..q_{l})\beta}\bigg),
\end{align*} where $f_v(\alpha)$ and $g(\alpha_1,\alpha_2)$
are given by
\begin{align*}
f_v(\alpha) & = (-1)^{v(e-r(g-1))}v(e-r(g-1)),\qquad v=1,2\\ \nonumber
g(\alpha_1,\alpha_2) & = (-1)^{e_1-e_2-(r_1-r_2)(g-1)}(e_1-e_2-(r_1-r_2)(g-1))\ \ .
\end{align*}

Comparing the coefficients of $\hat{g}_{\alpha}$ and using the rank one refined wall-crossing formula introduced in \cite{Chuang:2010ii}, we obtain the rank two refined wall-crossing formula: 
\begin{align} \label{wcf2}
&\mathrm{DT}_-(Q,2;y)=\sum_{\substack{q'\geq0,\ l \geq0,\ q_i>0 \\ q'+q_1+\cdots+q_l=Q}} \mathrm{DT}_+(q',2;y)\frac{1}{l!}\prod_{i=1}\kappa(f_2(q_{i}\beta)))\mathrm{DT}(q_i\beta;y)\\\nonumber&+
\sum_{\substack{q_1' > q_2'\geq 0\\ l\geq0,\ q_i>0 \\ q_1'+q_2'+ q_1+\cdots + q_l=Q}}  \frac{\kappa(g(q_1\beta,q_2\beta))}{2(y-y^{-1})}\mathrm{DT}_+(q'_1,1;y)\mathrm{DT}_+(q'_2,1;y)\frac{1}{l!}\prod_{i=1}^{l}\kappa(f_2(q_{i}\beta))\mathrm{DT}(q_i\beta;y)
\\\nonumber&-\sum_{ \substack{ q_2 > q_1\geq 0\\ q_1+q_2=Q\\ l \geq 0,\ \tilde{l} \geq 0\\q_1' \geq 0,\ q_2' \geq 0\\ n_i>0, \tilde{n}_i >0 \\ q_1'+n_1+\cdots+n_l = q_1 \\
q_2'+\tilde{n}_1+\cdots+\tilde{n}_{\tilde{l}} = q_2 }}\frac{\kappa(g(q_1\beta,q_2\beta)}{2(y-y^{-1})}\mathrm{DT}_+(q'_1,1;y)\mathrm{DT}_+(q'_2,1;y)\frac{1}{l!}\prod_{i=1}^{l}\kappa(f_1(n_{i}\beta))\mathrm{DT}(n_i\beta,y^k) \\ &\nonumber\ \ \ \ \ \ \ \ \ \ \ \ \ \ \ \ \ \ \ \ \ \ \ \times \frac{1}{\tilde{l}!}\prod_{i=1}^{\tilde{l}}\kappa(f_1(\tilde{n}_{i}\beta))\mathrm{DT}(\tilde{n}_i\beta;y^k).
\end{align}
As a consistency check the refined wall-crossing formula \eqref{wcf2} is specialized to \cite[Theorem 1.1]{Chuang:2010wx} and served as a refined generalization of the wall-crossing formula. 

\section{Higher D6 rank refined partition function of the local rational curve}
In this section we consider the local rational curve $\mathcal{O}(d_1)\oplus\mathcal{O}(d_2)\rightarrow\Sigma_0$ with $(d_1,d_2)=(-1,-1),(-2,0)$. We are going to apply \eqref{wcf2} to this case and derive the refined partition function with higher D6 rank. The computation on the local rational curve is possible since there exist a chamber in the moduli space such that the only BPS states are D6 state and the D2/D0 bound states \cite{Jafferis:2008uf}. This fact is not so obvious from the viewpoint of ADHM theory and is discussed in \cite[Section 5]{Chuang:2010wx} 

We denote $\gamma=(r,e,v)$ for the D2, D0 and D6 charge respectively. The intersection number for two charge vectors $\gamma'$ and $\gamma''$ is given by
\begin{equation*}
\langle\gamma',\gamma''\rangle= v''e'-v'e''+(v''r'-v'r'').
\end{equation*}
By the same arguments as in the corollary 5.5  and remark 5.6 in \cite{Chuang:2010wx}, the rational Donaldson-Thomas invariants in this geometry are
\begin{equation}\label{1}
\mathrm{DT}(r,e,0;y) = \left\{\begin{array}{ll}
{(-1)^{d_1-1}\over r[r]_y} & \mathrm{if}\ e=rn, \ n\in \mathbb{Z},\\
& \\
0 & {\mathrm{otherwise}}. \\ \end{array}\right. 
\end{equation}
\begin{equation} \label{eq:purevinv} 
\mathrm{DT}_\delta(0,0,1;y)=1,\qquad \mathrm{DT}_\delta(0,0,2;y)=\frac{1}{2[2]_{y}}.
\end{equation}
By the refined multicover formula \eqref{3}, we have the corresponding integer rational Donaldson-Thomas invariant 
$\overline{\mathrm{DT}}(1,n,0;y^k)=(-1)^{d_1-1}$.
The refined KS formula reads
\begin{equation}\label{eq:emptytoinftyA} 
\begin{aligned} 
\prod_{(r,n, v) \in \mathbb{Z}_{\geq 1}\times\mathbb{Z}_{\geq 0}\times \{0,1,2\} \cup \{0,0,1\} }^{\mu_{\delta=0} \uparrow}
\hat{U}_{\hat{e}_{(r,n,v)}}^{(\delta=0)} = 
\prod_{(r,n, v) \in \mathbb{Z}_{\geq 1}\times \mathbb{Z}_{\geq 0}\times \{0,1,2\} \cup \{0,0,1\}  }^{\mu_{\delta=\infty} \uparrow}
\hat{U}_{\hat{e}_{(r,n,v)}}^{(\delta=\infty)}.
\end{aligned} 
\end{equation}
The factors in each term are ordered in increasing order of $\delta_{\pm}$-slopes from left to right. For a sufficient large stability parameter $\delta_+$, the order of the  generator is the same as the unrefined invariants. Hence we have 
\[
e< {\delta_+ \over r} < \cdots <{e+\delta_+\over r} < {\delta_+\over r-1} < \cdots 
< {e+\delta_+\over r-1} < \cdots < \delta_+ + e < {2\delta_+ \over r} < \cdots < 
2\delta_+ + e. 
\]
The refined wall-crossing formula \eqref{eq:emptytoinftyA} then becomes
\begin{align*}
\nonumber&\exp\big((\frac{1}{y-y^{-1}}(\hat{f}_{00}+\frac{1}{2[2]_{y}}\hat{g}_{00})\big)\prod^{\infty}_{n=0}\hat{U}_{\hat{e}_{1,n}}\\
&=\prod_{n=0}^{e}\hat{U}_{\hat{e}_{1,n}}\prod_{n=0}^{e}\hat{U}^{(\infty)}_{\hat{f}_{r,n}}\prod_{n=0}^{e}\hat{U}^{(\infty)}_{\hat{f}_{r-1,n}}\cdots\prod_{n=0}^{e}\hat{U}^{(\infty)}_{\hat{f}_{1,n}}
\prod_{n=0}^{e}\hat{U}^{(\infty)}_{\hat{g}_{r,n}}\prod_{n=0}^{e}\hat{U}^{(\infty)}_{\hat{g}_{r-1,n}}\cdots\prod_{n=0}^{e}\hat{U}^{(\infty)}_{\hat{g}_{1,n}}\prod_{n=0}^{e}\hat{U}^{(\infty)}_{\hat{f}_{0,0}}
\end{align*}
where
\begin{align*}
\prod_{n=0}^e\hat{U}_{\hat{e}_{rn}}&=\prod_{n=0}^e\prod_{s\in \mathbb{Z}}\mathrm{E}(y^{s}\hat{e}_{r,n})^{-(-1)^{s}\Omega_{s}(r,n,0)}=\exp\sum_{ 0\leq n\leq e,\ k\geq1}\frac{\overline{\mathrm{DT}}_{\infty}(1,n,0;y^k)}{k(y^k-y^{-k})}\hat{e}_{kr,kn}\\
&=\exp\sum_{ 0\leq n\leq e,\ k\geq1}\frac{(-1)^{(d_1-1)}}{k(y^k-y^{-k})}\hat{e}_{kr,kn}=\exp\hat{\mathcal{H}}_{\hat{e}},\\
\prod_{n=0}^{e}\hat{U}^{(\infty)}_{\hat{f}_{rn}}
&=\prod_{n=0}^{e}\prod_{s\in \mathbb{Z}}\mathrm{E}(y^s\hat{f}_{r,n})^{-(-1)^{s}\Omega^{\infty}_{s}(r,n,1)}=\prod_{n=0}^{e}\prod_{s\in \mathbb{Z}}\prod_{i=0}^{\infty}\big(1+(-y)^{2i+1}y^s\hat{f}_{r,n}\big)^{(-1)^{s}\Omega^{\infty}_{s}(r,n,1)}\\&=\prod_{n=0}^{e}\exp\bigg(\frac{1}{y-y^{-1}}\overline{\mathrm{DT}}_{\infty}(n,r,1;y)\hat{f}_{n,r}+\frac{1}{y-y^{-1}}\overline{\mathrm{DT}}_{\infty}(n,r,1;y^2)\frac{1}{2[2]_{y}}\hat{g}_{2n,2r}\bigg)\\&=\exp\bigg(\frac{1}{y-y^{-1}}\sum_{n=0}^{e}\overline{\mathrm{DT}}_{\infty}(n,r,1;y)\hat{f}_{n,r}+\frac{1}{y-y^{-1}}\sum_{n=0}^{e}\frac{1}{2[2]_{y}}\overline{\mathrm{DT}}_{\infty}(n,r,1;y^2)\hat{g}_{2n,2r}\\&+\frac{1}{(y-y^{-1})^2}\sum_{\substack{n_2>n_1\geq 1\\ n_1+n_2\leq e}}\overline{\mathrm{DT}}_{\infty}(n_1,r_1,1;y)\overline{\mathrm{DT}}_{\infty}(n_2,r_2,1;y)
\\
&\times\kappa(n_1-n_2+r_1-r_2)\hat{g}_{n_1+n_2,r_1+r_2}\bigg),
\\
\prod_{n=0}^{e}\hat{U}^{(\infty)}_{\hat{g}_{rn}}&=\prod_{n=0}^{e}\prod_{s\in \mathbb{Z}}\mathrm{E}(y^s\hat{g}_{r,n})^{-(-1)^{s}\Omega^{\infty}_{s}(r,n,2)}=\prod_{n=0}^{e}\prod_{s\in \mathbb{Z}}\prod^{\infty}_{i=0}\big(1+(-y)^{2i+1}y^s\hat{g}_{r,n}\big)^{(-1)^{s}\Omega^{\infty}_{s}(r,n,2)}\\
&=\prod_{n=0}^{e}\exp\bigg(\frac{1}{y-y^{-1}}\overline{\mathrm{DT}}_{\infty}(n,r,1;y)\hat{g}_{n,r}\bigg).\\
\end{align*}
Multiplying both sides by the factor $\big(\prod_{n=0}\hat{U}_{\hat{e}_{1n}}\big)^{-1}$ and expanding the RHS, we obtain
\begin{align} \label{ge}
&\big(\prod_{n=0}^{e}\hat{U}_{\hat{e}_{1,n}}\big)^{-1}\exp(\frac{1}{y-y^{-1}}(\hat{f}_{00}+\frac{1}{2[2]_{y}}\hat{g}_{00}))\prod_{n=0}^{\infty}\hat{U}_{\hat{e}_{1,n}}=
\\&\mathrm{exp} \Big(\mathop{\frac{1}{y-y^{-1}}\sum_{1\leq s\leq r,\ 0\leq n \leq e}}
 {\mathrm{DT}}_{\infty}(s,n,1;y)\, { \hat{f}}_{sn}+ 
\mathop{\frac{1}{y-y^{-1}}\sum_{ 1\leq s\leq r,\ 0\leq n\leq e}}_{}{\mathrm{DT}}_{\infty}(s,n,2;y^2)\, 
{ \hat{g}}_{sn} +\nonumber \\
& \frac{1}{2}\frac{1}{y-y^{-1}}\sum_{\substack{r_1>r_2\geq 1, \ r_1+r_2\leq r,\ 
n_1,\ n_2\geq 0, n_1+n_2\leq e \\ \mathrm{or}  \ 
1\leq r_1=r_2\leq r/2, \ 0\leq n_1< n_2, \ n_1+n_2\leq e \\ \mathrm{or}  \ 1 \leq r_1 \leq r, \ 0 \leq n_1 \leq
e, \ r_2=n_2=0}}   \frac{(-y)^{(n_1-n_2+r_1-r_2)}-(-y)^{-(n_1-n_2+r_1-r_2)}}{y-y^{-1}} \nonumber\\
& {\mathrm{DT}}_{\infty}(r_1,n_1,1;y){\mathrm{DT}}_{\infty}(r_2,n_2,1;y)\,{ \hat{g}}_{r_1+r_2,n_1+n_2}+\cdots\Big).\nonumber 
\end{align}
Applying the BCH formula to the left hand side of equation \eqref{ge}, we have
\begin{align*}
&\big(\prod_{n=0}^{e}\hat{U}_{\hat{e}_{1n}}\big)^{-1}\exp\big(\frac{1}{y-y^{-1}}(\hat{f}_{00}+\frac{1}{2[2]_{y}}\hat{g}_{00})\big)\prod_{n=0}^{\infty}\hat{U}_{\hat{e}_{1n}}\\&=\exp\bigg(\frac{1}{y-y^{-1}}\big(\hat{f}_{00}+\frac{1}{2[2]_{y}}\hat{g}_{00}+\sum_{j=1}^{\infty}{1\over j!} [\underbrace{-\hat{\mathcal{H}}_{\hat{e}}, \cdots [ -\hat{\mathcal{H}}_{\hat{e}}}_{\text{j times}},\hat{f}_{00}+\frac{1}{2[2]_{y}}\hat{g}_{00} ]\cdots ]\big)\bigg),
\end{align*}
Next applying the following Lie algebra commutators,
\[
\begin{aligned} 
& [\hat{e}_{r_1,n_1}, \hat{f}_{r_2,n_2}] = \big( (-y)^{n_1+r_1}-(-y)^{-n_1-r_1}\big)\, \hat{f}_{r_1+r_2,n_1+n_2}\\
& [\hat{e}_{r_1,n_1}, \hat{g}_{r_2,n_2}] = \big( (-y)^{2n_1+2r_1}-(-y)^{-2n_1-2r_1} \big) \, \hat{g}_{r_1+r_2,n_1+n_2},\\
\end{aligned} 
\]
yields
\begin{align}\label{rank1}
&[\underbrace{-\hat{\mathcal{H}}_{\hat{e}}, \cdots [ -\hat{\mathcal{H}}_{\hat{e}}}_{\text{j times}},\hat{f}_{00} ]\cdots ]=\\&\nonumber\mathop{\sum_{n_1,\ldots, n_j = 0}^{e}}_{} 
\mathop{\sum_{k_1,\ldots, k_j \geq 1}}_{} 
(-1)^{j(d_1-1)}\prod_{i=1}^j  \frac{(-y)^{(n_i+1)k_i}-(-y)^{-(n_i+1)k_i}}{k_i(y^{k_i}-y^{-k_i})}(-1)^{j} \, \hat{f}_{k_1+\cdots + k_j, 
k_1n_1+\cdots+ k_jn_j},
\end{align}
and
\begin{align}\label{rank2}
&[\underbrace{-\hat{\mathcal{H}}_{\hat{e}}, \cdots [ -\hat{\mathcal{H}}_{\hat{e}}}_{\text{j times}},\frac{1}{2[2]_{y}}\hat{g}_{00} ]\cdots ]=\\&\nonumber\mathop{\sum_{n_1,\ldots, n_j = 0}^{e}}_{} 
\mathop{\sum_{k_1,\ldots, k_j \geq 1}}_{} 
(-1)^{j(d_1-1)}\prod_{i=1}^j  \frac{y^{2(n_i+1)k_i}-y^{-2(n_i+1)k_i}}{k_i(y^{k_i}-y^{-k_i})}(-1)^{j} \, \frac{1}{2[2]_{y}}{\hat{g}}_{k_1+\cdots + k_j, 
k_1n_1+\cdots + k_jn_j},
\end{align}
Define the refined partition functions for the refined Donaldson-Thomas invariants of $v$ units of D6 branes to be
\begin{equation}
Z_{v}^{\infty}(u,q,y)=\sum_{r\geq1}\sum_{n\in\mathbb{Z}}u^{r}q^{n}\mathrm{DT}_{\infty}(r,n-r,v;y).
\end{equation}
Replacing the generators in \eqref{rank1} and \eqref{rank2} by the monomial $u^{k}q^{kn}$, then the rank one refined partition function is given by
\begin{align} \label{rk1}
&\sum_{j=1}^{\infty}{1\over j!}\bigg(\sum_{n_i=0}^{e}\sum_{k_i}(-1)^{(d_1-1)j}\prod_{i=1}^{j}\frac{(-y)^{(n_i+1)k_i}-(-y)^{-(n_i+1)k_i}}{k_i(y^{k_i}-y^{-k_i})}\\&\nonumber\times(-1)^{j}u^{k_1+\cdots +k_j}q^{k_1n_1\cdots +k_jn_j + k_jn_j+k_1+\cdots + k_j}\bigg)
\\&\nonumber=\prod_{n=1}^{e+1}\prod_{t=0}^{n-1}\big(1-y^{-n+2t+1}u(-q)^{n}\big)^{(-1)^{(d_1-1)}} \ .
\end{align}
As a consistency check the rank one refined partition function becomes identical to the rank one refined partition function derived in \cite{Dimofte:2009bv}, after substituting  $d_1=-1$, $q_1=q/y$, $q_2=qy$ and $u=Q$ into \eqref{rk1}.

The rank two refined partition function of D6-D2-D0 on the local rational curve is
\begin{align*}
Z_{2}^{\infty}(u,q,y)=& \frac{1}{2[2]_{y}}\prod_{n=1}^{e+1}\prod_{t=0}^{2n-1}(1-y^{-2n+2t+1}uq^{n})^{(-1)^{(d_1-1)}}
- \sum_{\substack{r_1>r_2\geq 1, \ r_1+r_2\leq r,\ 
n_1,\ n_2\geq 0, n_1+n_2\leq e \\ \mathrm{or}  \ 
1\leq r_1=r_2\leq r/2, \ 0\leq n_1< n_2, \ n_1+n_2\leq e 
\\ \mathrm{or}  \ 1 \leq r_1 \leq r, \ 0 \leq n_1 \leq
e, \ r_2=n_2=0}} \\ \nonumber 
&  \frac{\kappa(n_1+r_1-n_2-r_2)}{2(y-y^{-1})}  
\mathrm{DT}_{\infty}(r_1,n_1,1,y)
\mathrm{DT}_{\infty}(r_2,n_2,1,y) q^{r_1+r_2}u^{n_1+n_2}.
\end{align*}
The substitution $y=1$ again give us back the Corollary 1.2 in \cite{Chuang:2010wx}.

\section{Higher D4 rank refined partition function of the local rational curve}
In this section, we study the resolved conifold $\mathcal{O}(-1)\oplus\mathcal{O}(-1)\rightarrow\Sigma_{0}$, with two D4 branes wrapping four cycle $\mathcal{O}(-1)\rightarrow \Sigma_{0} $ and the refined partition function of D4-D2-D0 bound states on them \cite{Nishinaka:2010fh}. We denote the D-brane charges as
\begin{equation}
\Gamma^{(w)}_{mn}=w\mathcal{D}+m\beta-ndV.
\end{equation} 
Where $\mathcal{D}\in H^2(X,\mathbb{R}),\beta\in H^4(X,\mathbb{R}),dV\in H^6(X,\mathbb{R})$ and $b,m,n$ denote the $D4,D2,D0$ brane charges. Taking the K{\"a}hler modulus $z$ of $\mathbb{P}^1$ from $\mathrm{Im}z>0$ to $\mathrm{Im}z<0$ corresponds to the flop transition of the resolved conifold. In the region $\mathrm{Im}z<0$ the four cycle is topologically $\mathbb{C}^2$ without compact four cycles, only D4 and D0 bound states exist.
As we vary the K{\"a}hler parameter, the walls of marginal stability can be identified. As the discussion in \cite{Aganagic:2012si}, the relevant wall of marginal stability in this case is the decay $\Gamma\rightarrow \Gamma_1+\Gamma_2$, where $\Gamma_2$ has only non-zero D2/D0 charge $(\pm1,N)$. We denote by $\overline{\mathrm{DT}}_{\pm}(m,n,w;y^k)$ the refined integer Donaldson-Thomas invariant of brane charges $\Gamma^{(w)}_{mn}$.
For convenience we introduce the following notations: $\hat{e}_{m,n}=\hat{e}_{(m,n,0)}$, $\hat{f}_{m,n}=\hat{e}_{(m,n,1)}$ and $\hat{g}_{\alpha}=\hat{e}_{(m,n,2)}$ and list the relevant factors in refined KS formula,
\begin{align*}
\hat{U}_{\hat{e}_{\pm1,N}}&=\prod_{s}\mathbf{E}(y^{s}\hat{e}_{\pm1,N})^{-(-1)^{s}\Omega_{s}(\pm1,N,0)}
=\exp\bigg(\sum_{k=1}^{\infty}\frac{\overline{\mathrm{DT}}(\pm1,N,0,y^k)}{k(y^k-y^{-k})}\hat{e}_{\pm1,N}\bigg),
\\
\hat{U}^{\pm\infty}_{\hat{f}_{m,n}}&=\prod_{s}\mathbf{E}(y^{s}\hat{f}_{m,n})^{-(-1)^{s}\Omega_{s}(m,n,1)}
\\\nonumber&=\exp\bigg(\frac{1}{y-y^{-1}}\overline{\mathrm{DT}}_{\pm}(m,n,1;y)\hat{f}_{m,n}+\frac{1}{y-y^{-1}}\overline{\mathrm{DT}}_{\infty}(m,n,1;y^2)\frac{1}{2[2]_y}\hat{g}_{m,n}\bigg),
\\
\hat{U}^{\pm\infty}_{\hat{g}_{m,n}}&=\prod_{s}\mathbf{E}(y^{s}\hat{g}_{m,n})^{-(-1)^{s}\Omega_{s}(m,n,2)}
=\exp\bigg(\frac{1}{y-y^{-1}}\sum_{q\geq 0 }\overline{\mathrm{DT}}_{\pm}(m,n,2;y)\hat{g}_{m,n,2}\bigg).
\end{align*}
We apply the wall-crossing formula \eqref{01}. We first expand the exponentials in $\hat{U}$ and then collect terms involving two D4 brane charges. Then we obtain the following equation
\begin{align} \label{wcf}
&\sum_{mn}\overline{\mathrm{DT}}_{-\infty}(m,n,1;y)\hat{g}_{m,n}+\sum_{m,n}\frac{1}{2[2]_y}\overline{\mathrm{DT}}_{-\infty}(m,n,1;y)\hat{g}_{2m,2n}
\\\nonumber&-\frac{1}{2}\sum[m_1-m_2]_{-y}\overline{\mathrm{DT}}_{-}(m_1,n_1,1;y)\overline{\mathrm{DT}}_{-\infty}(m_2,n_2,1,y)\hat{g}_{m_1+m_2,n_1+n_2}\\
\nonumber&=\prod_{N=0}^{\infty}\hat{U}_{\hat{e}_{\pm1,N}}\bigg\{\sum_{m,n}\overline{\mathrm{DT}}_{+\infty}(m,n,2;y)\hat{g}_{m,n}+\sum_{mn}\frac{1}{2[2]_y}\overline{\mathrm{DT}}_{+\infty}(m,n,1;y^2)\hat{g}_{m,n}\nonumber\\&-\frac{1}{2}\sum_{m_1>m_2}[m_1-m_2]_{-y}\overline{\mathrm{DT}}_{+\infty}(m_1,n_1,1;y)\overline{\mathrm{DT}}_{+\infty}(m_2,n_2,1;y)\hat{g}_{m_1+m_2,n_1+n_2}
\bigg\}\prod_{N=0}^{\infty}\hat{U}^{-1}_{\hat{e}_{\pm1,N}} \ . \nonumber
\end{align}
Denote the generating function of the refined Donaldson-Thomas invariant by
\begin{equation*}
Z_{w}^{\pm\infty}(q,u)=\sum_{m,n}\mathrm{DT}_{\pm\infty}(m,n,w;y)q^mu^n \ . 
\end{equation*}
We replace the generators by the monomial $u^mq^n$ for the D0-D2 charges. The equation $\eqref{wcf}$ for two D4 branes then becomes
\begin{align*}
& Z_{2}^{-\infty}(q,u,y)+\frac{\kappa(m_1-m_2)}{y-y^{-1}}\mathrm{DT}_{-\infty}(m_1,n_1,1;y)\mathrm{DT}_{-\infty}(m_2,n_2,1;y)u^{m_1+m_2}q^{m_1+m_2}\\&=\prod_{n=0}^{\infty}\prod_{t=0}^{1}(1+y^{2t-1}uq^n)^{-1}\prod_{n=1}^{\infty}\prod_{t=0}^{1}(1+y^{2t-1}u^{-1}q^n)^{-1}\times\bigg( Z_{2}^{+\infty}(u,q,y)\\&-\frac{1}{2}\sum_{m_1>m_2}[m_1-m_2]_{-y}\mathrm{DT}_{+\infty}(m_1,n_1,;y)\mathrm{DT}_{+\infty}(m_2,n_2,1;y)q^{m_1+m_2}u^{n_1+n_2}\bigg).
\end{align*}
Since the intersection number of the charges with one non-compact D4-brane in this configuration is one, the generating function is identical to the unrefined one derived in \cite{Nishinaka:2010qk}
\begin{align*}
&Z_{1}^{+\infty}(u,q,y)=f(q,y)\sum_{n\in\mathbb{Z}}(-1)^nq^{\frac{n(n-1)}{2}}u^{n},\\
&Z_{1}^{-\infty}(u,q,y)=f(q,y)\prod_{n=1}^{\infty}(1-q^n).
\end{align*}
The prefactor $f(q,y)$ is related to the D0 and D4 brane bound state which cannot be determined since the D4 branes are non-compact. The remaining part of the partition function $Z_{1}^{+\infty}(u,q,y)$ is related to the D2 branes bound to D0 and D4 branes which are on the compact genus zero curve $\Sigma_0$. Therefore, the $\overline{\mathrm{DT}}_{-}(m,n,1,y)$ with non-zero $m$ is zero and the second term of left hand side vanishes.
\begin{align*}
& Z_{2}^{-\infty}(u,q,y)=\prod_{n=0}^{\infty}\prod_{t=0}^{1}(1+y^{2t-1}uq^n)^{-1}\prod_{n=1}^{\infty}\prod_{t=0}^{1}(1+y^{2t-1}u^{-1}q^n)^{-1}\\
&\times\bigg( Z_{2}^{+\infty}(u,q,y)-\frac{1}{2}[f(q,y)]^{2}\sum_{m_1>m_2}[m_1-m_2]_{-y}q^{\frac{m_1(m_1-1)}{2}+\frac{m_2(m_2-1)}{2}}(-u)^{m_1+m_2}\bigg).
\end{align*}
Using the argument in \cite{Nishinaka:2010qk}, we can conclude that the D4 branes wrap the whole fibre and localize on the $P^1$ in the large radius limit $\mathrm{Im}z\rightarrow-\infty$. Thus the generating function $Z_{2}^{-\infty}(u,q,y)$ is independent of D2 brane charge. The refined partition function of two D4-branes in the limit $\mathrm{Im}z\rightarrow\pm\infty$ is given by
\begin{align} \label{twoD4}
&Z_{2}^{+\infty}(u,q,y)=[f(q,y)]^2\prod_{n=1}^{\infty}(1-q^n)^2 \prod_{n=0}^{\infty}\prod_{t=0}^{1}(1+y^{2t-1}uq^n)\prod_{n=1}^{\infty}\prod_{t=0}^{1}(1+y^{2t-1}u^{-1}q^n)\\&+\frac{1}{2}[f(q,y)]^{2}\sum_{m_1>m_2}[m_1-m_2]_{-y}q^{\frac{m_1(m_1-1)}{2}+\frac{m_2(m_2-1)}{2}}(-u)^{m_1+m_2} \ , \nonumber \\
&Z_{2}^{-\infty}(u,q,y)=[f(q,y)]^2\prod_{n=1}^{\infty}(1-q^n)^2 \ . 
\end{align}
After substituting $q=e^{\phi_0}$ and $uq^{-1/2}=e^{\phi_1}$ into \eqref{twoD4} and using the Jacobi triple product formula, the first term of $Z_{2}^{+\infty}(u,q,y)$ is identical to the partition function of the rank two $(q,t)$-deformed Yang-Mills theory on $\mathcal{O}(-1)\rightarrow \Sigma_{0}$ derived in \cite{Aganagic:2012si},
\begin{equation*}
Z^{(q,t)YM}(\phi_0,\phi_1,y)=[f(\phi_0,y)]^2\sum_{n_i\in\mathbb{Z}^2}e^{-\frac{1}{2}\phi_0n^2}e^{-\phi_1\sum_{i}n_i}y^{\sum_i(3-2i)n_i},
\end{equation*}
 where $\phi_0$ and $\phi_1$ are the D0 and D2 brane chemical potentials related to the gauge  parameters $g_s$ and $\theta$. The second term indicates that there are two-centered bouned states in the limit $\mathrm{Im} z\rightarrow +\infty$ shown by supergravity analysis \cite{Nishinaka:2010fh}. 

\section{Membranes and sheaves}
In this section we discuss how the computations in the current paper fit into the conjectural correspondence between the enumerative geometry of curves in a Calabi-Yau 5-fold $Z$ and the 1-dimensional sheaves on the 3-folds $X$ embedded in $Z$, which arise as the fixed loci of a $\mathbb{C}^{\times}_q$-actions on $Z$ \cite{Nekrasov:2014nea}. Here we recall the conjecture and some theorems assuming the conjecture. 

Let $Z$ be a Calabi-Yau 5-fold, admitting a $\mathbb{C}^{\times}_q$-action such that 3-folds $X$ in $Z$ are the fixed points of the action. $X$ could have multiple components. 
Consider the stable pairs on $X$ consisting of a pair $(\mathcal{F},s)$, where $\mathcal{F}$ is a pure 1-dimensional sheaf and $s$ is a section of $\mathcal{F}$ such that the cokernal of $s$ is zero-dimensional.
The condition imposed on the pair $(\mathcal{F},s)$ is called PT stability condition. Indeed such a condition can be recast in terms of the polynomial stability condition \cite{PBSC} and is related to the Donaldson-Thomas theory by a wall-crossing in the polynomial stability condition space.

Let $\text{PT}(X)$ be the moduli stack of the PT pairs on $X$. The morphism $\Pi_{\text{PT}}:\text{PT}(X) \to \text{Chow}(X)$ is the Hilbert-Chow morphism, constructed by taking the scheme-theoretical support of the sheaf $\mathcal{F}$ in the pair. Conjecturally there exist an membrane moduli  $\text{M2}(Z)$, describing certain types of membrane configurations wrapping 2-cycles in $X$, together with a similar Hilbert-Chow morphism $\Pi_{\text{M2}}:\text{M2}(Z) \to \text{Chow}(X)$. Then we have the following diagram:

\begin{equation}
\xymatrix{
\text{M2}(Z) & \text{M2}(Z)^{\mathbb{C}^{\times}_q} \ar@{->}[l]_{\displaystyle{\iota}} \ar@{->}[rd]_{\displaystyle{\Pi_{\text{M2}}}}
&& \text{PT}(X) 
\ar@{->}[ld]^{\displaystyle{\Pi_{\text{PT}}}}\\
&& \text{Chow}(X)
}
\end{equation}
where $\iota$ is the inclusion of the fixed locus of the $\mathbb{C}^{\times}_q$-action. Both $\text{M2}(Z)$ and $\text{PT}(X)$ are virtually smooth. We denote their virtual structure sheaves by $\OO_{\text{M2}}$ and $\OO_{\text{PT}}$ respectively. 

Assuming an equivariant virtaul localization theorem can be proven for $\OO_{\text{M2}}$, we have
\begin{align}
\iota_* \tilde{\OO}_{\text{M2, localized}} &=  \tilde{\OO}_{\text{M2}} \ , \\ \nonumber
\tilde{\OO}_{\text{M2, localized}} &=  \iota^{-1} \tilde{\OO}_{\text{M2}} \ , 
\end{align}
in the equivariant $K$-theory of ${\text{M2}}(Z)$ and $\text{M2}(Z)^{\mathbb{C}^{\times}_q}$. 
$\tilde{\OO}_{\text{M2}}$ is the modified virtual structure sheaf, whose precise definition will be omitted here. 

We define the natural $S(d)$-invariant maps 
\begin{equation*}
\Sigma_d: \text{Chow}(X)^{\times d} \to \text{Chow}(X)
\end{equation*} by 
\begin{equation*}
\left( C_1, \dots , C_d\right) \to \sum C_i \,.
\end{equation*}
Given a sheaf $\mathcal{F}$ on $\text{Chow}(X)$, we define the symmetric algebra over $\text{Chow}(X)$ 
\begin{equation}
S_{\text{Chow}} \ \mathcal{F}  = \bigoplus_{d=0}^{\infty} \left( \Sigma_{d,*} \mathcal{F}^{\boxtimes d} \right)^{S(d)} \ . 
\end{equation}

One of the conjectures in \cite{Nekrasov:2014nea} states the following:

\begin{conjecture}\cite[Conjecture 1] {Nekrasov:2014nea} \label{conj}
Let $G_q$ be the centralizer of $\mathbb{C}^{\times}_q$ in $\text{Aut}(Z, \Omega^5)$ and $T(Z)= H^2(Z, \mathbb{C}) / ( 2 \pi i H^2(Z, \mathbb{Z})/\text{torsion})$. Then we have the following equality in $T(Z) \times G_q$--equivariant $K$-theory of the $\text{Chow}(X)$: 
\begin{equation}
S_{\text{Chow}}\,  \Pi_{\text{M2},*} \ \left( \iota_*^{-1} \ \tilde{\OO}_{\text{M2}} \right) = \Pi_{\text{PT},*} \ \left(
\tilde{\OO}_{\text{PT}} \otimes \Phi \right) \label{conj1}
\end{equation}
where $\Phi$ is some explicit computable combination of the universal sheaves
on $\prod \text{PT}(X_i)$ describing the interaction of the components of $X$.
\end{conjecture}

Let $Z_r$ be the $A_{r-1}$-surface fibration over a Calabi-Yau 3-fold $X$, described explicitly in \cite[3.2.1]{Nekrasov:2014nea}. On $X$, instead of the rank one PT stable pair theory $(\mathcal{F}, s)$, we can consider the rank $r$ PT pair with $r$ sections $(s_1, \cdots, s_r)$, 
\begin{equation*}
\OO_X^r \stackrel{\oplus s_i}{\longrightarrow} \mathcal{F}
\end{equation*} with the same stability condition, {\it i.e.} $\mathcal{F}$ is pure 1-dimensional and the cokernel of the complex is zero-dimensional.

Assuming Conjecture \ref{conj}, it is proved in \cite[Section 5.4]{Nekrasov:2014nea} that the LHS of the conjecture computes exactly the rank $r$ Donaldson-Thomas/PT invariants on $X$, after the interaction term $\Phi$ is taken care of.   
Namely in this situation $( \tilde{\OO}_{\text{PT}} \otimes \Phi )$ on the RHS of  Conjecture \ref{conj} should be replaced by a new modified virtual structure sheaf $\tilde{\OO}_{\text{PT},r}$ on the PT moduli. Therefore it is natural to conjecture that what we compute in this paper is the generating function of the equivariant Euler characteristics of  $\tilde{\OO}_{\text{PT},r}$ with $r=2$ and $X$ being local curve geometries. 

\section{Discussion and conclusion}
In this paper we present some computations of higher rank refined Donaldson-Thomas invariants on local curve geometries, corresponding to local D6-D2-D0 or D4-D2-D0 configurations. A refined wall-crossing formula for invariants with higher D6 or D4 ranks is derived and verified to agree with the existing formulas under the unrefined limit. Using the formula, refined invariants on the $(-1,-1)$ and $(-2,0)$ local rational curve with higher D6 ranks are computed. We also use the formula to compute the partition function of two D4 branes wrapping on the $\mathcal{O}(-1)\rightarrow \Sigma_{0}$ and give the refined extension of the result \cite{Nishinaka:2010fh}. 

In the large radius limit, the rank 2 formulas for D4-D2-D0 systems give rise to the partition function of $(q,t)$-deformed Yang-Mills theory with two-centered bound states terms. 
The generalization of the refined invariants with arbitrary ranks can be recursively determined by the lower rank invariants and it should give the partition function of rank $N$ $(q,t)$-deformed Yang-Mills theory with many multi-centered bound states terms. The whole computation, although tractable in principle, can be foreseen to be quite lengthy and is omitted here.

The higher rank refined Donaldson-Thomas invariants on other local toric geometries such as $\mathbb{C}^3/Z_n$ and $\mathbb{C}^3/Z_2 \times Z_2$ should also have the similar structures. Since the stability conditions and the chamber structures are much more complicated in these cases, applying refined wall-crossing formulas is probably not the most efficient way to obtain the higher rank refined invariants.
So in the end we would like to mention that there has been related work by Gholampour-Kool-Young \cite{GKY}, on the rank 2 invariants on toric 3-folds. And it would be interesting to extend their work to treat the higher rank refined invariants of toric 3-folds.

\bibliography{ref}
\bibliographystyle{abbrv}

\end{document}